\documentclass{iau-JDSS}

\title{Scaling relations in early-type galaxies from integral-field stellar kinematics}

\author[Cappellari et al.]{M.\ Cappellari,$^1$ N.\ Scott,$^1$ K.\ Alatalo,$^2$ L.\ Blitz,$^2$ M.\ Bois,$^3$ F.\ Bournaud,$^4$ M.\ Bureau,$^1$ R.\ L.\ Davies,$^1$ T.\ A.\ Davis,$^1$ P.\ T.\ de~Zeeuw,$^{5,14}$ E.\ Emsellem,$^{5,3}$ J.\ Falcon-Barroso,$^6$ S.\ Khochfar,$^7$ D.\ Krajnovic,$^1$ H.\ Kuntschner,$^5$ P.-Y.\ Lablanche,$^3$ R.\ M.\ McDermid,$^8$ R.\ Morganti,$^9$ T. Naab,$^{10}$ M.\ Sarzi,$^{11}$ P.\ Serra,$^9$ R.\ C.\ E.\ van den Bosch,$^{12}$ G.\ van~de~Ven,$^{13}$ A.\ Weijmans$^{14}$ and L.\ M.\ Young$^{15}$}

\affiliation{
$^1$University of Oxford, UK;
$^2$University of California, Berkeley, USA;
$^3$Universit\'e de Lyon, France;
$^4$Universit\'e Paris Diderot, France;
$^5$ESO, Garching, Germany;
$^6$IAC, La Laguna, Spain;
$^7$MPI for Extraterrestrial Physics, Garching, Germany;
$^8$Gemini Observatory, Hilo, USA;
$^9$ASTRON, Dwingeloo, The Netherlands;
$^{10}$Universit\"ats-Sternwarte M\"unchen, Germany;
$^{11}$University of Hertfordshire, Hatfield, UK;
$^{12}$The University of Texas, Austin, USA;
$^{13}$IAS, Princeton, USA;
$^{14}$Leiden University, The Netherlands;
$^{15}$New Mexico Tech, Socorro, USA
}

\pubyear{2009}
\volume{Volume 15}
\jname{Highlights of Astronomy, Volume 15}
\editors{Ian F.\ Corbett, ed.}
\begin{document}

\maketitle

Early-type galaxies (ETGs) satisfy a now classic scaling relation $R_{\rm e}\propto\sigma_{\rm e}^{1.2} I_{\rm e}^{-0.8}$, the Fundamental Plane (FP; Djorgovski \& Davis 1987; Dressler et al.\ 1987), between their size, stellar velocity dispersion and mean surface brightness. A significant effort has been devoted in the past twenty years to try to understand why the coefficients of the relation are not the ones predicted by the virial theorem $R_{\rm e}\propto\sigma_{\rm e}^{2} I_{\rm e}^{-1}$.

Recent studies, using independent approaches from either (i) detailed dynamical models or (ii) strong galaxy lensing, point to a genuine variation of the mass-to-light ratio $M/L$ in galaxies as the reason for nearly all the observed `tilt' in the FP (e.g.\ Cappellari et al.\ 2006; Bolton et al.\ 2008). However these studies are limited by a small and biased sample or are restricted to only the most massive ETGs respectively.

We overcome both limitations by modeling the stellar dynamics, using axisymmetric Jeans anisotropic models (JAM; Cappellari 2008), for the $K$-band selected, volume-limited ATLAS$^{3D}$ sample of 263 nearby ETGs, spanning a large range of masses and with $60<\sigma_{\rm e}<350$ km s$^{-1}$. A key for the project is the availability for all galaxies of high-quality integral-field kinematics observed with the SAURON spectrograph and detailed Multi-Gaussian Expansion (Emsellem et al. 1994) models of the photometry.

We confirm the genuine $M/L$ variation and construct both the FP and the More FP (MFP; Bolton et al.\ 2007) for the ATLAS$^{3D}$ sample, relating the mean surface density $\Sigma_{\rm e}\equiv I_{\rm e}\times(M/L)_{\rm JAM}$, $\sigma_{\rm e}$ and $R_{\rm e}$. Our MFP produces a relation as tight as the FP over the full mass range. We compare the global $(M/L)_{\rm JAM}$ variation among galaxies with predictions from two-SSP stellar population models and find that variations of both dark matter (or IMF) and population are required to explain the observations.

\end{document}